\documentstyle[prl,aps,twocolumn]{revtex}

\begin{document}
\draft
\title{Stimulated Enhancement of Cross-Section by a Bose-Einstein Condensate}
\author{J.J. Hope \cite{email:Hope} and C.M. Savage}
\address{Department of Physics and Theoretical Physics,
The Australian National University, \\
Australian Capital Territory 0200, Australia.}
\date{\today}
\maketitle

\begin{abstract}
This paper examines the feasibility of constructing an experiment 
which detects the atomic stimulation of a photon emission process.  A 
beam of atoms (bosons) in an excited state is passed over an atomic 
trap which traps the atoms when they are in their internal ground 
state.  When the trap contains a Bose-Einstein condensate, the 
cross-section for absorption of the atomic beam is increased.  We 
examine a particular model in which this {\it atom}-stimulation is 
observable, and is also characterised by the emission of photons in a 
narrow cone in the direction of the atomic beam.
\end{abstract}


\narrowtext

A large number of atoms in a single quantum mechanical (internal and 
external) state has been produced in atomic traps in the recent 
experiments which have produced a Bose-Einstein condensate
(BEC)\cite{Rb,Li,Na}.  This high degree of quantum degeneracy may be 
used as an atomic source for atom optics experiments
\cite{Holland,Wiseman,Martin,Guzman}, or to demonstrate new effects 
which depend entirely on the quantum statistics.  The stimulation of a 
transition by a BEC has not been observed.  We propose a method by 
which the emission of photons may be observed to be stimulated by 
the presence of a BEC.

The emission of a photon from an atom in an excited state can be 
stimulated by a large number of photons in the destination mode, as in 
a laser.  If the atoms involved are bosons, then the same emission can 
also be stimulated by the presence of many atoms in the destination 
{\it atomic} mode.  Any transition rate between two states is enhanced 
by a factor of $(N+1)$ where $N$ is the number of bosons occupying the 
final state.

An atom in a metastable excited state can pass through an atomic trap 
which only traps the ground (internal) state.  If the trap contains a 
condensate of $N$ atoms, then there will be an enhancement of the 
fraction of the spontaneous emission which goes into the ground (trap) 
state, and therefore an enhancement of the overall emission rate.  
A schematic of this process is shown in figure 1.  If we let $\gamma$ 
be the free space spontaneous emission rate and $f$ be the fraction of 
the atoms which will spontaneously emit into the ground (trap) state, 
then the total emission rate $\gamma_{tot}$ is the sum of the 
spontaneous emission rate into the non-ground states and the 
spontaneous plus the stimulated emission rates into the ground state

\begin{equation}
	\gamma_{tot} = \gamma (1-f) +(N+1)\gamma f = \gamma+ \gamma N f.
	\label{EmissionRate}
\end{equation}

The rightmost term of this equation will be called the {\it stimulated} 
emission rate, referring to the photon emission which is stimulated by 
a highly populated {\it atomic} mode.  It is clear from this equation 
that for a BEC with a sufficiently large number of atoms, the 
stimulated emission rate will dominate the spontaneous emission rate, 
and that the fraction $f$ will determine the critical size of such a 
BEC.  In this letter, we will calculate the ratio of stimulated 
emission to spontaneous emission for a particular model, and show that 
for experimentally realisable parameters it is possible for the 
stimulated emission rate to become much larger than the spontaneous 
emission rate, significantly increasing the overall emission rate.  
In this particular model, the stimulated emission will occur in a 
narrow cone around the direction of motion of the excited atoms.

We denote the center of mass wavefunction of an excited (internal) 
state atom in the beam by $\Psi({\bf x},t)$.  If we also denote the 
single particle ground (internal and trap) state by $\Phi({\bf x})$, 
and the momentum kick produced by the photon emission by $\hbar {\bf 
k}$, then the fraction $f$ of atoms which will spontaneously emit into 
the ground (internal and trap) state is given by the overlap integral

\begin{equation}
	f(t) = \int_{\Omega}\frac{d^2{\bf k}}{4\pi k_o^2}\left\|\int{
	\bf dx}\Phi^*({\bf x})\Psi({\bf x},t) e^{i{\bf k}.{\bf 
	x}}\right\|^{2},
	\label{eq:ftime}
\end{equation}
where $\hbar k_{o}$ is the absolute momentum kick given by the 
emission of the photon and $\int_{\Omega}d^2{\bf k}$ denotes the 
integral in k-space of all possible photon directions.

We now assume that the excited state wavefunction will be travelling 
with some narrow momentum distribution centred around $(\hbar 
k_{o},0,0)$, so that there will be a strong resonance in the integral 
around ${\bf k}=(-k_{o},0,0)$.  This means that we can ignore all 
contributions to the k-space integral except those very close to this 
point, and the photons will emit in a very narrow cone in the 
direction of travel of the excited atom.  This means that the 
transverse (y,z) integrals will be trivial, and we will approximate 
them to be unity.  This approximation is valid if the transverse 
wavefunction of the excited atom has the same shape as the transverse 
wavefunction of the destination state.  The overlap integral is then 
approximately one dimensional

\begin{equation}
	f(t) = \int_{-k_{o}}^{k_{o}}\frac{dk}{2 k_{o}}\left\|\int_{-\infty
	}^{\infty}dx \Phi^*(x)\Psi(x,t) e^{ik.x}\right\|^{2}.
	\label{eq:frac1D}
\end{equation}

We now consider a specific case to calculate this overlap integral, 
and determine the feasibility of producing an experiment which will 
measure {\it atom}-stimulated emission.  We consider the case where 
both wavefunctions have a Gaussian form, and the excited state 
wavefunction $\Psi(x,t)$ is shifted to a centre of momentum of $\hbar 
k_{o}$

\begin{eqnarray}  \label{eq:gstate}
	\Phi(x) & = & (\frac{1}{2 \pi l_{1}^{2}})^{\frac{1}{4}} 
	\mbox{exp}\left[-\frac{x^{2}}{4l_{1}^{2}}\right],\\
	\label{eq:estate}
	\nonumber
	\Psi(x,t) & = & (\frac{1}{2 \pi l_{2}^{2}s(t)})^{\frac{1}{4}} 
	\mbox{exp}\left[i(k_{o}x-\omega_{o}t)\right] \times\\
	&&\rule{20mm}{0mm}\mbox{exp}\left[-(x-v_{o}t)^{2}/(4l_{2}^
	     {2}s(t))\right],
\end{eqnarray}
where $l_{1}$ and $l_{2}$ are the sizes of the Gaussian wavefunctions 
of the ground state and the excited state wavefunctions respectively, 
$\omega_{o}=\hbar k_{o}^{2}/(2 M)$, $v_{o}=\hbar k_{o}/M$, $s(t)=1+i 
\hbar t/(2Ml_{2}^{2})$ is a measure of the spreading of the beam, 
and $M$ is the mass of the atom.  We have assumed here that the source 
atomic beam and the ground state of the trap have been perfectly 
aligned, and we have set the zero of time at the moment when they are 
exactly coincident.

These wavefunctions are chosen to be Gaussian because that is the form 
of the ground state of a harmonic trap.  A possible method of 
producing the excited state atoms in this form would be to produce a 
BEC in a second trap, and then use a laser pulse to transfer the atoms 
to a metastable state via a Raman transition.  A Raman transition 
would leave the atoms in a metastable internal state with an overall 
momentum kick $\hbar k_{o}$ due to the difference in photon momenta.  
This method would be appropriate when the kick was required to be 
small.  We will show that the largest number of atoms are drawn into 
the BEC when a {\it small} kick is given to the atoms.  This type of 
atom source has several natural advantages for this experiment.  
Firstly, the BEC is a very cold atom source and will allow all of the 
atoms to be in the energy range which can be trapped by the target 
BEC.  Secondly, using copropagating lasers to produce the Raman 
transition will provide naturally the required resonance between the 
motional state of the atom (Gaussian wavepacket travelling with 
average momentum $\hbar k_{o}$) and the kick ($\hbar k_{o}$) due to 
the photon which may be emitted.

We now make the assumption that the excited state wavepacket does not 
spread while it overlaps the target wavefunction $\Phi(x)$ ($s(t)=1$), 
or precisely that $\Psi(x,t)=\Psi(x-v_{o}t,0)$.  This approximation is 
made for calculational purposes, and is true for timescales $\tau\ll 
2Ml_{2}^{2}/\hbar$.  This implies the condition

\begin{equation}
	k_{o}\gg \sqrt{l_{1}^{2}+l_{2}^{2}}/(2l_{2}^{2}).
	\label{spreadCondition}
\end{equation}

We now calculate the fraction $f(t)$ of atoms spontaneously emitting 
into the ground state under these assumptions.  From Eqs.\ 
(\ref{eq:frac1D},\ref{eq:gstate},\ref{eq:estate}) we obtain the 
result:

\begin{eqnarray}  
	\nonumber
	f(t) &=&  \frac{\sqrt{\pi}}{2k_{o}\sqrt{2(l_{1}^{2}+l_{2}^{2})}} 
	 \mbox{ Erf}\left[\frac{2\sqrt{2}k_{o}l_{1}l_{2}}
	{\sqrt{l_{1}^{2}+l_{2}^{2}}}\right] \times\\
	\label{fresult}
	&&\rule{15mm}{0mm}\mbox{exp}\left[-(v_{o}t)^{2}/(2(l_{1}^
	      {2}+l_{2}^{2}))\right].	
\end{eqnarray}

In general, the excited atoms will have a very weak spontaneous 
emission rate, which is due to the low frequency of the emitted 
photon.  This means that we can ignore depletion of the excited state 
atoms when calculating the number of spontaneous emission events.  We 
also ignore depletion when calculating the total number of stimulated 
emissions.  Including depletion is a straightforward extension, and 
can only become important after the stimulated emission already has 
increased the total number of emission events by a large factor.

In an experiment, it is likely that the atoms will travel some 
distance $D$ between the source and the detectors which is large 
compared to the size of the trap or the atomic beam, $l_{1}$, $l_{2}$.  
The atoms will be emitting spontaneously for the entire distance, but 
the stimulated emission will be insignificant unless they are 
overlapping with the trap ground state.  Integrating each of the two 
terms of the right hand side of equation (\ref{EmissionRate}) with 
respect to time will allow us to calculate the total number of 
stimulated emission events and the total number of spontaneous 
emission events.  We denote the ratio of these by $R$, which is given 
by

\begin{equation}   \label{eq:ratio}
	R = \frac{N \pi}{2 D k_{o}}
    \mbox{ Erf}\left[\frac{2\sqrt{2}k_{o}l_{1}l_{2}}
    {\sqrt{l_{1}^{2}+l_{2}^{2}}}\right].
\end{equation}

The dependence of this ratio on $N$ was expected, as the stimulated 
emission rate is directly proportional to the number of atoms in the 
BEC so the total number of stimulated emissions will be proportional 
to $N$.  The total number of spontaneous emissions is proportional to 
the time of travel of the excited atomic cloud.  This in turn is 
proportional to the distance $D$, so we expected the ratio $R$ to be 
inversely related to $D$.  The dependance on $k_{o}$ is a consequence 
of our particular model.

Figure 2 plots the value of $R$ as a function of $k_{o}$, and shows that 
smaller values of $k_{o}$ produce a stronger stimulated signal.  The 
other restriction on $k_{o}$ is provided by the calculational 
requirement that the excited state wavepacket doesn't spread over 
time.  Over the region of the trap, this will be true if inequality 
(\ref{spreadCondition}) is satisfied.

Substituting some realistic numbers into this equation will show the 
feasibility of conducting such an experiment.  We choose the number of 
atoms in the condensate to be $N=5\times 10^{5}$, which has been 
achieved already in experiments \cite{Na}.  The size of both the 
incident wavepacket and the target ground state are 
$l_{1}=l_{2}=30\mu$m, which is a couple of times larger than current 
experiments.  The total distance travelled by the excited atoms is 
chosen to be $D=2$cm, but this should be as small as possible in a 
well designed experiment, and in the theoretical limit can be as small 
as the traps themselves.  For a value of $k_{o}=4 \times 
10^{6}\mbox{m}^{-1}$, the excited state wavepacket increases in size 
by about half a percent while passing over the ground state of the 
trap.  These parameters give a ratio of $R=10$ times more stimulated 
emission events than spontaneous emission events.  If the spontaneous 
emission is very weak but still measurable, then there will be a 
significant increase in the emission rate when the BEC is present in 
the trap.

There are two possible methods for detecting the emission of photons.  
The first is to directly measure the proportion of excited state atoms 
which manage to pass through the trap.  For weak spontaneous emission 
but strong stimulated emission, this will change as the BEC absorbs 
the atomic beam.  Another method of detection which can be carried out 
simultaneously is to detect the photons which are emitted.  
Spontaneously emitted photons will come from the condensate at a 
random direction, whereas the radiation due to the stimulated process 
will emit in a narrow cone in the direction of travel of the excited 
state atoms.  For the parameters used in the previous calculation, the 
maximum deviation from the center of the cone would be of the order of 
$9^{o}$, corresponding to a solid angle of $0.006 \times 4\pi$.  This 
angle is smaller for larger values of $k_{o}$, $l_{1}$ and $l_{2}$.  
The background spontaneous emission rate over this small solid angle 
will therefore be correspondingly reduced, increasing the signal to 
noise ratio by several orders of magnitude.

If we consider the effect of collecting the light over a small solid 
angle, we can arrange it so that nearly all of the stimulated photons 
are collected while only a small proportion of the spontaneously 
emitted photons are collected.  We estimate the size of this relative 
increase by estimating the size of the k-space resonance and dividing 
it by the total area of the k-sphere.  The k-space resonance of the 
stimulated signal for small solid angles can be derived from equations 
(\ref{eq:frac1D},\ref{eq:gstate},\ref{eq:estate}), which show that it 
has a width of approximately $\sqrt{1/l_{1}^{2}+1/l_{2}^{2}}$.  This 
gives us an approximation to the ratio $R_{p}$ of photons produced by 
stimulated and spontaneous emission within the reduced solid angle

\begin{equation}   \label{eq:rgamma}
	R_{p} \approx \frac{N \pi}{D \sqrt{1/l_{1}^{2} + 1/l_{2}^{2}}}
    \mbox{ Erf}\left[\frac{2\sqrt{2}k_{o}l_{1}l_{2}}
    {\sqrt{l_{1}^{2}+l_{2}^{2}}}\right].
\end{equation}

We can see that for this detection scheme, if we have $k_{o}$ large 
enough to ensure a small solid angle (required by our original 
approximations), then the error function will be unity, and the result 
will be independent of $k_{o}$, depending only on the size of the 
trapped state and the excited state wavefunction.  This detection 
scheme may allow the use of a single photon transition to produce the 
source beam while still measuring the effect of atom-stimulation, but 
for large $k_{o}$ the depletion of the excited state beam will be 
insignificant.

Gravity has not been considered in this model, but will be important 
over the timescales involved with this experiment.  It may be possible 
to utilise gravity in a particular experimental arrangement, but it 
may be necessary to balance the gravitational force.  This might be 
achieved with a far-detuned light force \cite{HopeRaman} or an atom 
waveguide such as a hollow optical fiber
\cite{Savage,Marksteiner,Ito}.

Superradiance will not enhance the background spontaneous emission 
rate provided that the {\it source} of atoms is sufficiently dilute 
that on average, the atoms are much further apart than one optical 
wavelength.  This condition can be easily satisfied, as there is no 
fundamental theoretical reason to use a high density source, and the 
stimulated effect has been shown in equation (\ref{eq:ratio}) to be 
largest when the wavepacket of the atomic {\it source} is made as 
physically large as possible.  We also assume that the effect of 
collisions between the excited state atoms and the ground state atoms 
is negligible.  This is true provided the {\it target} BEC is 
sufficiently dilute.  This is a restriction on the density of the 
target BEC, but we have also shown that the stimulated emission rate 
is maximal when the {\it target} ground state wavefunction is 
physically as large as possible.

Reabsorption of the emitted photons by the condensate will reduce the 
spontaneous and stimulated emission signals by an equal 
proportion\cite{Morice}.  This effect will not alter the signal to 
noise ratio, but it would tend to heat the target BEC.  We have 
neglected the effects of reabsorption of the beam.

This paper has shown that it should be feasible to produce an 
experiment which directly measures a process which is stimulated by a 
large number of bosonic atoms in a single quantum mechanical state.  
If these experiments can be produced such that the stimulated emission 
completely dominates the spontaneous emission, and a significant 
number of atoms are transferred into the BEC, then this may be a 
method of increasing the number of atoms in a BEC beyond that made 
possible by evaporative cooling.

The authors would like to thank A. White for his stimulating 
discussion.

\begin{figure}
\caption{Schematic of the modelled system.  The top wavepacket shows 
the beam of atoms in the excited state.  The other smooth solid line 
shows the trap potential for the ground electronic state of the atom.  
The dashed lines show the energy levels in the trap, with a large 
population in the the ground trap state.  The waved lines show 
transitions into various trap states.}
\end{figure}

\begin{figure}
\caption{Ratio ($R$) of stimulated emission to spontaneous emission.  
The chosen parameters are $N=5\times 10^{5}$, $D=10$cm and $k_{o}$ is 
in units of $(l_1^2+l_2^2)^{1/2}/(l_1 l_2)$.}
\end{figure}

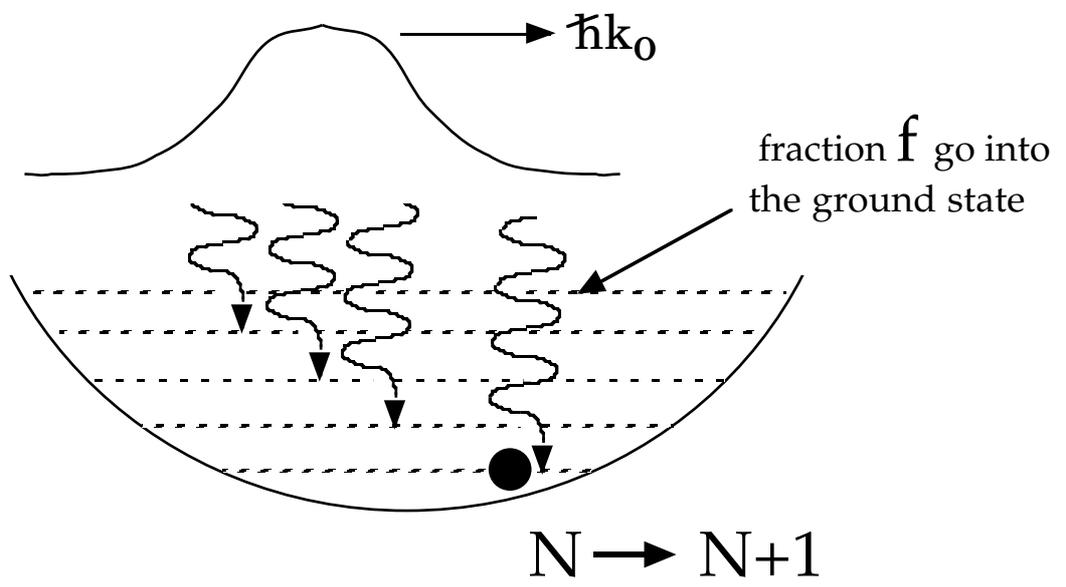
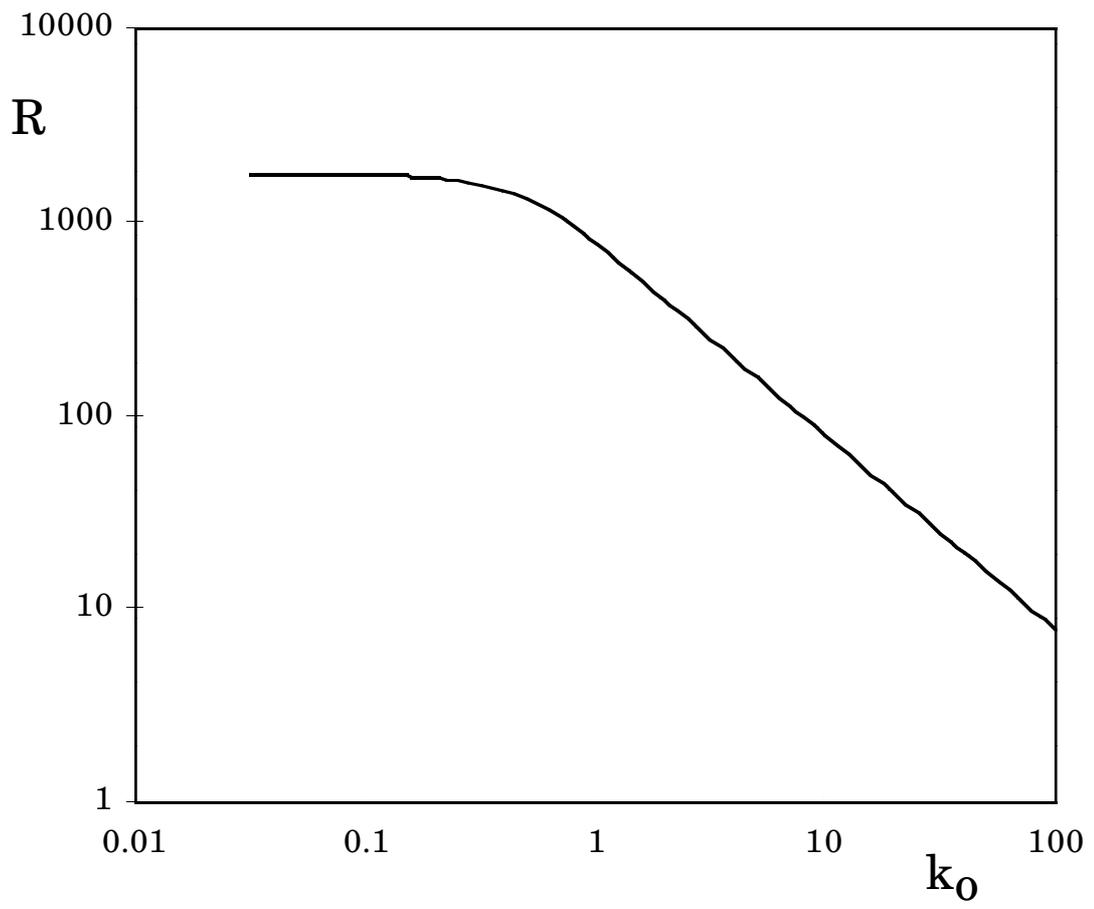
\end{document}